\documentclass[12pt]{article}
\usepackage{amsmath}
\usepackage{amssymb}
\usepackage{amsfonts}
\usepackage{amscd}
\usepackage{amsbsy}
\usepackage{amsthm}
\usepackage{latexsym}%%
\usepackage{slashed}

\def\nn{\nonumber}       %%%    nonumber
\def\n{\label}                 %%%    nonumber
\def\r{\ref}                    %%%    nonumber

\def\beq{\begin{eqnarray}}
\def\eeq{\end{eqnarray}}
\def\ln{\,\mbox{ln}\,}

\def\Det{\,\mbox{Det}\,}

\def\Tr{\,\mbox{Tr}\,}
\def\sTr{\,\mbox{sTr}\,}

\def\al{\alpha}
\def\be{\beta}

\def\ga{\gamma}
\def\de{\delta}
\def\vp{\varepsilon}
\def\ep{\epsilon}

\def\la{\lambda}
\def\na{\nabla}
\def\pa{\partial}

\def\si{\sigma}

\def\ph{\varphi}
\def\ta{\tau}

\def\Ga{\Gamma}

\def\La{\Lambda}

\begin{document}

%%%%%%%%%%%%%%%%%%%%%%%%%%%%%%
\begin{center}

{\Large \bf Renormalization of Yukawa model with sterile scalar in curved
space-time}

\vskip 8mm
 {\bf V\'{\i}tor Fernandes Barra}$^{a}$\footnote{E-mail
address: \ vitorbarra@ice.ufjf.br}, \  \ {\bf Iosif L.
Buchbinder}$^{b,c}$\footnote{E-mail address: \ joseph@tspu.edu.ru},
\  \ {\bf Jarme Gomes}$^{a}$\footnote{E-mail address: \
jarme@ice.ufjf.br},

{\bf Andreza Rairis Rodrigues}$^{a}$\footnote{
E-mail address: \ andrezarodrigues@ice.ufjf.br},
\  \
{\bf Ilya L. Shapiro}$^{a,b,c}$\footnote{E-mail address: \ shapiro@fisica.ufjf.br}
\vskip 4mm

$^a$ \
{\sl
Departamento de F\'{\i}sica, \ ICE, \ Universidade Federal de Juiz de Fora,
36036-330 Juiz de Fora, \ MG, \ Brazil}
\vskip 2mm

$^b$ \
{\sl
Department of Theoretical Physics, Tomsk State Pedagogical
University,  634061 Tomsk, Russia}
\vskip 2mm

$^c$ \
{\sl
National Research Tomsk State University, 634050 Tomsk, Russia}
\end{center}

%%%%%%%%%%%%%%%%%%%%%%%%%%%
\begin{quotation}
\noindent
{\bf Abstract.}
\ \ 
We explore the classical and quantum properties of a sterile
scalar field coupled to $N$ copies of Dirac fermions in an external
gravitational field.  We find that the self-interaction scalar potential
of a model that is consistent at the quantum level, includes odd (first
and third) powers of a scalar.  In particular, one has to consider,
besides the standard non-minimal coupling of the form $\,\xi \ph^2 R$,
the new type of non-minimal coupling of the form $\,f\ph R\,$ with
new non-minimal parameter $f$. We study the one-loop renormalization
of such a theory including renormalization of the new non-minimal
coupling. Also, we calculate the one-loop effective potential using
the renormalization group and show how the renormalization group
analysis should be extended compared to the standard expression
which was derived in 1980-ies. This conclusion is supported by the
direct calculation of effective potential using normal coordinates and
covariant cut-off regularization. The important features of the classical
theory with a sterile scalar are related to the presence of the qualitatively
new terms in the induced action of gravity, coming from the odd terms.
We show that this new feature of the theory may have phenomenologically
relevant consequences, both in the low-energy gravitational physics and
at the high energies, corresponding to inflation.
\vskip 2mm

%%%%%%%%%%%%%%%%%%%%%%%%%%
\vskip 2mm

%%%%%%%%%%%%%%%%%%%%%%%%%%
\end{quotation}

%%%%%%%%%%%%%%%%%%%%%%%%%%%%%%
%%%%%%%%%%%%%%%%%%%%%%%%%%%%%%
%%%%%%%%%%%%%%%%%%%%%%%%%%%%%%
\section{Introduction}
\label{int}

The unique scalar field of the Minimal Standard Model (MSM) of
particle physics is the Higgs boson, which is complex and belongs to
the fundamental representation of $SU(2)$. The extensions of MSM
such as nonminimal, supersymmetric version, of Grand Unification
Theories, typically have larger scalar sector, but the new scalars are
always representations of the symmetry group of the corresponding
particle physics model. The consistency of such models with respect
to quantum field theory requirements are the main tool in restricting
the extensions of MSM, and in particular the scalar sector.

At the same time there is  another sort of scalar fields, which are
intensively used in cosmology. Both inflaton and quintessence are
real scalars which are not related to representations of the symmetry
group of particle physics and which can be called sterile scalars. An
interesting question concerns the restrictions which can be imposed
in quantum theory on the self-interaction potential of such a scalar
field. A practical realization of this program requires assuming the
form of the interaction between the sterile scalar and elementary
particles.
In the present work we consider the simplest possible version of such
interaction, that means we consider a sterile scalar coupled to the
$N$ copies of massive Dirac fermion through Yukawa interaction.

It is well known that the multiplicative renormalization of a scalar
field in curved space-time requires introducing the non-minimal
coupling between scalar field and gravity in the form $\xi \ph^2 R$.
However, if the classical potential of the scalar field includes
$\ph^3$-term, one can expect that the renormalized theory should
include the new type of the non-minimal coupling proportional to
$\ph R$, with the new nonminimal parameter. Thus, we arrive at the
problem of describing the quantum structure of the theory with a
sterile scalar. The purpose of this paper is to consider the main
aspects of this problem, such as renormalization and renormalization
group. As far as we know, these subjects have not been studied in
the literature, so far.

To study the above new aspects we start by using the standard heat
kernel approach to derive the one-loop divergences in such a model
in curved space-time. As a result of this calculation we arrive at the
minimal form of consistent self-interacting potential which provides
multiplicative renormalizability of the model. The main new feature
of this potential in comparison with a usual scalar (e.g. Higgs)
is the presence of third and first powers of scalar field in the
classical potential. Let us note
that in the previous existing calculations of similar divergences in
Refs.~\cite{BuchSha-90,book,TomsJHEP,TomsPRD} the presence of
these terms was recognized, but the consequences of this aspect of
the theory were never sufficiently well explored. We fill this gap in
the present work, including the discussion of the role of odd terms
for the induced action of gravity, when the scalar field is in the
minimum of the potential of the sterile scalar.
%%%%%%  red
%%% {\LARGE \textbullet}
Another interesting aspect of the sterile scalar coupled to fermions
through the Yukawa interaction is the possible role of the odd terms
is inflation, that will be also addressed in what follows.

From the formal QFT side, the challenging problem is how to take
into account the renormalization group for the odd terms, especially
when it concerns the renormalization group - based derivation of the
effective potential.
%%%   red
%%%   \textcolor{red}{\LARGE \textbullet}
The standard
expression for effective potential restored from the renormalization
group equation for effective action in curved space \cite{BuchOd84}
(see also further development of the renormalization group method,
applied for other sectors of effective action in \cite{BW})
is valid only for the theory where divergences have only second and
fourth powers of the scalar field. The effective potential in the model
with Yukawa interaction was calculated recently in
\cite{TomsJHEP,TomsPRD} for the special case of massless
fermions, when the odd terms in the scalar potential are not necessary
and the loop contributions to the potential can be derived on the basis
of the standard general expressions in flat \cite{ColeWein} and
curved spacetime \cite{BuchOd84,book}. If fermions are massive
and there are odd terms, these standard results are incomplete and
the renormalization group derivation of the effective potential
should be modified somehow. As usual
(see the discussion in \cite{CorPot}), the  renormalization group
derivation is based on the identification of the renormalization
parameter $\mu$, which enables one to easily go beyond the
local potential approximation \cite{BW,book}. On the other
hand, such identification represents an assumption, which is always
good to check, at least in the simplest case of effective potential.
Thus, in order to achieve an additional verification of the complete
result, we perform the derivation of effective potential for a sterile
scalar directly, using the method which was developed recently in
\cite{CorPot} in the basis of  Riemann normal coordinates and local
momentum representation \cite{BunchParker79,ParkerToms}.

The paper is organized as follows. In the next Sec.~2 we describe the
derivation of one-loop divergences in the model with a sterile scalar
coupled to $N$-component fermion. The beta- and gamma-functions
are calculated, and the one-loop effective potential restored from
the renormalization group in Sec.~3 . In Sec.~4 we present the
tree-level analysis of the curved-space analog of the spontaneous
symmetry breaking, in the presence of odd scalar terms, using an
approximation of small odd terms and weak gravitational field.
%%%%%%  red
%% {\LARGE \textbullet}
The analysis of the  induced gravitational action  in Sec.~5 shows
that in the presence of odd terms, for a light scalar case, there are
unusual low-energy terms, which can be phenomenologically
interesting. At another end of energy spectrum we meet small but
potentially detectable effects of the odd terms. Depending on the
details of the particle physics model at high energies, these terms
may be within the reach of the possible observations.
%% \textcolor{red}{\LARGE \textbullet}
Finally, in Sec.~6 we draw our conclusions and discuss
the perspectives for further work.

%%%%%%%%%%%%%%%%%%%%%%%%%%%%
%%%%%%%%%%%%%%%%%%%%%%%%%%%%
\section{Yukawa model with sterile scalar and one-loop divergences}
\label{sect2}

Consider Yukawa model with a single real sterile scalar coupled to
the $N$ copies of fermion field.  It proves useful to choose the
classical action of the form
\beq
S &=&
\int d^{4} x \sqrt{-g} \Big\{
i \bar{\Psi}_i  \left( \ga^{\mu} \na_{\mu}
+ iM + ih\varphi  \right)\delta^{ij}  \Psi_j
+ \frac12 \left( g^{\mu\nu}\pa_{\mu}\ph\pa_{\nu}\ph
- m^2 \ph^2 + \xi R \ph^{2} \right)
\nn
\\
&-&
\frac{\la}{4!} \ph^4 -\frac{g}{3!} \varphi^{3}
- \tau \varphi - fR \ph
\Big\},
\n{1}
\label{classact}
\eeq
where $m$ is a scalar fields mass, $M$ is a spinor field mass, $h$ is
the Yukawa coupling constant, $\la,$ $g$ and $\tau$  are the coupling
constants in the scalar sector which survive in the flat limit, while
$\xi$ and $f$  are the non-minimal parameters of scalar field
coupling to gravity. The terms with odd powers and correspondingly
the parameters  $g$,  $\tau$  and $f$ were not analysed in detail in
the previous considerations of the model in
\cite{BuchSha-90,book,TomsJHEP,TomsPRD}, regardless the odd
terms were identified in the one-loop divergences.
%%%%%%  red
%% {\LARGE \textbullet}
From the formal quantum field theory point of view, the consistent
theory of
a sterile scalar coupled to fermions should include these terms from
the very beginning and this is the approach we are starting to pursue
in this work.
%% \textcolor{red}{\LARGE \textbullet}

Regardless of the calculation of one-loop divergences in this theory
follows the standard procedure (see, e.g., \cite{book} for a number
of well-elaborated examples), we shall give some details below, in
order to simplify possible verifications.
Let us stress from the very beginning that gravity will not be
quantized in this paper (indeed, the generalization to the case of
quantum gravity can be found for similar models in the original
papers \cite{BuSh-HDQG} and book \cite{book}, where the
one-loop calculations were done by means of the generalized
Schwinger-deWitt technique, and the subsequent recent work
\cite{Agrav}, where similar calculations were performed by using
Feynman diagrams (without much detail) and used for attempting
to  reconciliate the renormalizability of quantum gravity and the
absence of higher derivative ghosts.

Let us start by decomposing the matter fields into classical $\,\ph$,
$\,\bar{\Psi}$, $\,\Psi$  and quantum $\si,\,\bar{\eta},\, \eta$
counterparts,
\beq
\ph \rightarrow \ph + \sigma ,
\qquad \bar{\Psi}_i \rightarrow \bar{\Psi}_i + \bar{\eta}_i,
 \qquad \Psi_j \rightarrow \Psi_j + \eta_j.
\eeq
The one-loop divergences are defined by the bilinear part of the
action, which involves the operator $ \hat{H}$,
\beq
S^{(2)} &=& \frac12 \int d^{4}x \sqrt{-g}\,\, \Big( \si \quad
\bar{\eta}_i \Big) \,\hat{H} \, \left(
\begin{array}{c}
\si  \\
\eta_j
\end{array}
\right)
\nn
\\
&=&
\frac{1}{2} \int d^{4} x \sqrt{-g} \Big\{ \sigma H_{11} \sigma
+ \bar{\eta}_i H_{21} \sigma + \sigma H_{12} \eta_j
+ \bar{\eta}_i H_{22} \eta_j \Big\} ,
\label{action}
\eeq
or in the explicit form
\beq
S^{(2)}
&=&
\frac{1}{2}\int d^{4} x \sqrt{-g} \Big\{ 2i \bar{\eta}_i
(\slashed{\nabla} + iM)\eta_j  \delta^{ij} - \sigma \Box \sigma
- m^{2} \sigma^{2} + \xi R \sigma^{2}
\nn
\\
&-&
2h(\ph \bar{\eta}_i \eta_j
+ \sigma \bar{\Psi}_i \eta_j + \sigma \bar{\eta}_i \Psi_j)\delta^{ij}
- \frac{\lambda}{2}\sigma^{2} \ph^{2} -g\ph \sigma ^2 \Big\}.
\n{bili}
\eeq

Here the quadratic in quantum fields action $S^{(2)}$ depends on the
background gravitational field and the background fields $\bar{\Psi}$,
$\Psi, \varphi$. After some algebra we get
%% \begin{strip}
\beq
\hat{H}
&=&
\left( \begin{array}{cc}
\xi R - \Box - m^{2}  -g \ph - \frac{\lambda}{2} \ph^{2} &
-2h \bar{\Psi}_j
\\
-2h \Psi_i & 2i(\slashed{\nabla} + iM + i h \ph) \delta^{ij}
\end{array}
\label{hatH}
\right).
\nn
\\
\eeq
%% \end{strip}
In order to reduce the problem of deriving $\ln\Det \hat{H}$ to the
standard form, one can introduce the conjugated matrix operator
$\hat{H}^{*}$ as follows,
\beq
\hat{H}^{*} = \left(
\begin{array}{cc}
-1 & 0 \\
0 & -\frac{1}{2} (i\slashed{\nabla} + M)
\end{array}
\right).
\eeq

It is well-known that the one-loop effective action has the form
$\sim \Tr\ln(\hat{H})$. To calculate the divergences of effective
action we will write it as
\beq
\Tr\ln(\hat{H}) \,=\,\Tr\ln(\hat{H}\hat{H}^*)-\Tr\ln(\hat{H}^*).
\eeq
It is easy to see that $\Tr\ln \hat{H}^*$ contributes only to the
vacuum  divergences, that are already known for an arbitrary model
\cite{birdav,book}. Therefore it is sufficient to calculate the
divergences of the product $\hat{H}\hat{H}^{*}$, which has a
standard form,
%% \begin{strip}
\beq
{\cal\hat{H}} &=&
\hat{H}\hat{H}^{*} = \hat{1} \Box  + 2 \hat{h}^{\mu} \nabla_{\mu} + \hat{\Pi},
\label{H}
\eeq
%% \end{strip}
\\
where we can identify
\beq
\hat{h}^{\mu} = \left(
\begin{array}{cc}
0 & \frac{i}{2} h \bar{\Psi}_j \gamma^{\mu} \\
0 & \frac{i}{2} h \ph \gamma^{\mu} \delta^{ij}
\end{array}
\right)
\label{hmu}
\eeq
and
\beq
\hat{\Pi} = \left(
\begin{array}{cc}
m^{2} + \frac{\lambda}{2} \ph^{2} - \xi R + g\ph &  hM \bar{\Psi}_j
\\
2h \Psi_i & \ \delta^{ij}\Big[ M^{2} -\frac{1}{4} R + hM \ph \Big]
\end{array}
\right).
% \n{eq: 11}
\nn
\\
\label{Pi}
\eeq

The one-loop divergences can be derived by means of the
Schwinger--De-Witt technique (see, e.g. \cite{book}, \cite{BV})
and are given by the general expression
\beq
\Ga^{(1)}_{div}
&=&
-\frac{\mu ^{D-4}}{\vp} \int d^{D}x \sqrt{-g}
\Tr \Big\{ \frac{1}{2}\hat{P}^{2} + \frac{1}{12}\hat{S}_{\mu
\nu}^{2}
+ \frac{1}{6} \Box \hat{P}
\nn
\\
&+&
\frac{\hat{1}}{180}
\big(R_{\mu \nu \alpha \beta}^{2} - R_{\mu \nu}^{2} + \Box R \big)
\Big\}, \mbox{\quad}
\eeq
where $\vp=(4\pi)^{2}(D-4)$ and
\beq
\hat{P}
&=&
\hat{\Pi} + \frac{\hat{1}}{6}R - \nabla_{\mu} \hat{h}^{\mu}
- \hat{h}_{\mu} \hat{h}^{\mu},
\nn
\\
\hat{S}_{\mu \nu}
&=&
\big[ \nabla_{\nu}, \nabla_{\mu} \big] \hat{1}
+ \nabla_{\nu} \hat{h}_{\mu} - \nabla_{\mu} \hat{h}_{\nu}
+ \hat{h}_{\nu} \hat{h}_{\mu} - \hat{h}_{\mu} \hat{h}_{\nu}. \quad \quad
\label{Gadivs}
\eeq

We give intermediate formulas in the Appendix, and here present
only the final result,
\beq
\Gamma^{(1)}_{div}
&=&
- \frac{\mu ^{D-4}}{\vp}
\int d^{D} x \sqrt{-g} \,\bigg\{
\frac{m^4}{2} - 2NM^4
+ \Big[ \frac{N}{3}M^2
- m^2\Big(\xi - \frac{1}{6} \Big) \Big]R
+ 2Nh^2 (\partial_\mu \ph)^2
\nn
\\
&+&
\frac{8N-1}{180} R_{\mu \nu} ^2
+ \Big[\frac{5N+1}{45} - \frac{1}{6}\Big(\xi
- \frac{1}{6}\Big)\Big]\Box R
+ \sum_k 3ih^2 \bar{\Psi}_k\Big[\frac{1}{2} \slashed{\nabla}
- iM -  ih \ph  \Big]\Psi_k
\nn
\\
&-&
\Big[\frac{1}{2}\Big(\xi - \frac{1}{6}\Big)^2 + \frac{N}{18}\Big]R^2
+ \frac{1}{6}\,\Big( g\ph - 8NhM \Big) \Box \ph
+  \frac{1}{2}\big(g^2 + \lambda m^2-24Nh^2 M^2\Big)\ph ^2
\nn
\\
&+&
 \frac{1}{12}\,\Big(\lambda - 16 Nh^2 \Big)\Box \ph^2
\label{Gadiv}
- \frac{1}{2} \Big[ \Big(\xi - \frac{1}{6}\Big) \la
-\frac{2}{3}N h^2 \Big]R \ph ^2
+ \Big( \frac{N}{24} +\frac{1}{45}\Big) R_{\mu \nu \alpha \beta} ^2
\nn
\\
&+& (m^2 g - 8NhM^3)\ph
+ \Big( \frac{1}{8} \lambda ^2 -2 N h^4 \Big) \ph ^4
- \Big( 8N M h^3 - \frac12\,g \lambda \Big)\ph ^3
\nn
\\
&-&
\Big[  g\Big(\xi - \frac{1}{6}\Big) -\frac{2}{3}N hM \Big] R\ph
\bigg\}, \qquad \qquad
\eeq
where the vacuum divergences were also included for completeness.

A few general comments are in order at this point. First of all, the
result (\ref{Gadiv}) confirms our expectations. All odd terms which
we included into the classical action (\ref{action}) really emerge in
the one-loop divergences. The reason is that in the theory with
sterile scalar these terms are not protected by any kind of symmetry,
and hence it was actually expected that they would show up. Second,
as far as we have the odd-power divergences, one should expect the
logarithmic contributions in the corresponding finite part of effective
action, in particular in the effective potential of a sterile scalar. In the
next sections we shall see that these expectation will be completely
confirmed. Third, it is worth pointing out that the odd terms may affect
on the form of the non-local form factors, similar to what we had in the
Yukawa model for the even terms \cite{danteYuk}
and earlier for a self-interacting scalar \cite{GBP+Gorbar}. The
discussion of this issue goes beyond the framework of the present
work and will be left for the future.

The renormalization relations between bare and renormalizable
quantities have the form which directly follows from the divergences.
For the fields we have
\beq
\ph_0
&=&
\mu ^{\frac{D-4}{2}} \left(1+ \frac{2Nh^2}{\epsilon}\right) \ph ,
\n{par:phi}
\\
\Psi_{k0}
&=&
\mu ^{\frac{D-4}{2}} \left(1+ \frac{3}{4\epsilon}h^2\right)\Psi_k ,
\\
\bar{\Psi}_{k0}
&=&
\mu ^{\frac{D-4}{2}} \left(1+ \frac{3}{4\epsilon} h^2\right)\bar{\Psi}_k .
\label{fields}
\eeq
The relations for masses have the form
\beq
M_0
&=&
\left(1-\frac{9}{2\epsilon}h^2\right) M ,
\\
m_0 ^2
&=&
m^2
- \frac{ g^2 +4N h^2 m^2  + \lambda m^2 -24N h^2 M^2}{\epsilon} .
\label{masses}
\eeq
For the usual even couplings and nonminimal parameters we have
\beq
\xi_0
&=&
\xi -  \frac{\la + 4Nh^2}{\epsilon} \, \Big(  \xi -\frac{1}{6}  \Big) ,
\\
h_0
&=&
\mu ^{\frac{4-D}{2}}
h\left(1 - \frac{4Nh^2 + 9h^2}{2\epsilon} \right) ,
\\
\lambda_0
&=&
\mu^{4-D}
\left( \lambda +\frac{48 N h^4 -8 N \la h^2 - 3\la^2 }{\ep} \right) .
\label{parameters1}
\eeq
And, finally,  for the odd couplings and nonminimal parameters,
\beq
g_0
&=&
\mu ^{\frac{4-D}{2}} \left(g + \frac{48N M h^3 - 3g\lambda
-6N h^2 g}{\epsilon} \right) ,
\\
\tau_0
&=&
\mu ^{\frac{D-4}{2}}\left(\tau
+ \frac{8N h M^3 -2N\tau h^2 - m^2 g }{\epsilon} \right) ,
\\
f_0
&=&
\mu ^{\frac{D-4}{2}} \left[ f+ \frac{g}{\epsilon}
\Big(\xi - \frac{1}{6}\Big)
- \frac{2N hM + 6N f h^2}{3\epsilon} \right].
\label{parameters2}
\eeq
These expressions demonstrate the non-trivial renormalization of the
odd coupling parameters, including the new non-minimal parameter $f$.

%%%%%%%%%%%%%%%%%%%%%%%%%%%%
%%%%%%%%%%%%%%%%%%%%%%%%%%%%
\section{Renormalization group and effective potential}
\label{sect3}

In this section we consider the renormalization group equation for
the effective potential and discuss its solution to derive the
effective potential for the model under consideration up to first
order in scalar curvature. The form of the equation is defined by
the corresponding beta-and gamma-functions which are calculated
on the basis of the renormalization relations for the parameters and
fields (for the theories in curved space time see e.g.
\cite{Buch84,Toms83,book}).

Let us begin with  beta-functions . They are defined as follows
\beq
\beta_{P} &=& \lim_{D \to 4} \mu \frac{dP}{d \mu},
\label{betas-def}
\eeq
where $P=\{m, M, h, \lambda, \xi, g, \tau, f\}$
are the renormalized parameters.
The scheme of derivation is described in \cite{book} and we will not
repeat it here, but only present the results. The calculation which is
based on the relations (\ref{masses}), (\ref{parameters1}),
(\ref{parameters2}) leads to
\beq
\beta_h
&=&
\frac{(4N + 9)h^3}{(4\pi)^2} ,
\nn
\\
\beta_M
&=&
\frac{9h^2 M}{2(4\pi)^2} ,
\nn
\\
\beta_{\lambda}
&=&
\frac{1}{(4\pi)^2} \Big(8N\lambda h^2 + 3\lambda ^2 - 48N h^4 \Big) ,
\nn
\\
\beta_\xi
&=&
\frac{1}{(4\pi)^2}
\Big(4Nh^2 + \lambda
\Big) \Big(\xi - \frac{1}{6}\Big)  ,
\nn
\\
\beta_g
&=&
\frac{1}{(4\pi)^2} \Big(\frac{3}{2} g\lambda + 3 N gh^2 - 12N Mh^3 \Big) ,
\nn
\\
\beta_{m^2}
&=&
\frac{1}{(4\pi)^2}
\Big[m^2 \lambda + g^2 + \Big(4 m^2 - 24M^2 \Big)Nh^2 \Big] ,
\nn
\\
\beta_{\tau}
&=&
\frac{1}{(4\pi)^2}\Big(2N \tau h^2 + g m^2 -8N hM^3 \Big) ,
\nn
\\
\beta_f
&=&
\frac{1}{(4\pi)^2}
\Big[ 2N f h^2 - g \Big(\xi - \frac{1}{6}\Big) + \frac{2}{3} N Mh\Big].
\label{betas}
\eeq

The gamma-functions are defined as follows:
\beq
\lim_{D \to 4} \,\,\mu \frac{d \Phi}{d \mu}\, = \,\gamma_{\Phi} \Phi,
\eeq
where $\Phi$ are the renormalized fields, $\Phi= (\ph , \,\Psi_k)$.
The relations (\ref{fields}) lead to
\beq
\gamma_\ph &=& -\frac{2N h^2}{(4\pi)^2},
\\
\gamma_{\Psi_{k}} &=& -\frac{3h^2}{4(4\pi)^2}.
\label{gammas}
\eeq

In the case of conformal invariant theory we should put all
dimensional constants $m^2$,  $M$,  $g$,  $\tau$ and $f$ vanish
and set $\xi=\frac16$. It is easy to see that in this situation the pole
coefficient in the expression for the divergences (\ref{Gadiv}) is
also conformal invariant. Furthermore, the beta functions have the
corresponding conformal fixed point, as it has to be from the
general perspective \cite{Buch84,book}.

Now we briefly discuss how to find the one-loop effective potential
from the
$\overline{\rm MS}$ renormalization group equation in curved
spacetime. The starting point is the overall $\mu$-independence
of effective action,
\beq
\mu\frac{d}{d\mu}\,\Ga[g_{\al\be},\Phi,P,n,\mu]\,=\,0,
\label{n8}
\eeq
which immediately leads to \cite{Buch84,book}
\beq
\Big\{\,\mu\frac{\pa}{\pa\mu}+\be_P
\,\frac{\pa}{\pa P}&+&\int d^Dx
\,\,\ga_{\Phi}\Phi \,\frac{\de}{\de \Phi (x)}
\,\Big\}
\times
\Ga[g_{\al\be},\Phi,P,D,\mu]=0,
\label{n12}
\eeq
where we assume the sum over all parameters $P$ and the fields
$\Phi=(\ph,\,\Psi_k)$. From now on we shall set $D=4$.

The effective potential is defined as zero-order approximation in
the derivative expansion for the scalar sector of $\Ga$,
\beq
\Ga[g_{\al\be},\ph,P,\mu]&=&
\int d^4x \sqrt{-g}\,\bigg\{ - V_{eff}\big(\ph,g_{\al\be}\big)
+ \frac{1}{2}\,Z\big(\ph,g_{\al\be}\big)g^{\mu\nu}
\pa_\mu \ph \pa_\nu \ph \,+\,\dots \bigg\}. \qquad
\label{effpot}
\eeq
Since (\ref{n12}) is a linear homogeneous equation, we get
\beq
\Big\{\,\mu\frac{\pa}{\pa\mu}+\be_P
\,\frac{\pa}{\pa P}
\,+\,\ga_\ph \,\ph\, \frac{\pa}{\pa \ph }
\,\Big\}
V_{eff}(g_{\al\be},\ph,P,\mu)=0.
\label{RGeffpot}
\eeq

Equation (\ref{RGeffpot}) means that the explicit functional
dependence on $\mu$ in the effective potential is exactly compensated
by the $\mu$-dependence of the scalar field $\ph$ and parameters $P$.
At the one-loop level the last dependence can be written in the simple
form involving the first order logarithmic dependence, as one can
figure out from the above expressions for beta- and gamma-functions.

We will search for the effective potential up to the terms linear in scalar
curvature, $V_{eff}=V_{0}+RV_{1}$, where $V_{0}$ is the flat-space
effective potential and $RV_{1}$ is the first curvature-dependent
correction to $V_{0}$. Both functions $V_{0}$ and $V_{1}$ satisfy
equation (\ref{RGeffpot}). Before solving the equations for $V_{0}$
and $V_{1}$, we will take into account that in the one-loop
approximation each spin gives additive contribution to the effective
action. Therefore we can write
$V_{0}=V^{(0)}_{0}+V^{(\frac{1}{2})}_{0}$
and
$V_{1}=V^{(0)}_{1}+ V^{(\frac{1}{2})}_{1},$
where the labels $(0)$ and $(\frac{1}{2})$  mean contribution from
quantum scalar and spinor fields, respectively. The quantities
$V^{(0)}_{0}$ and $V^{(\frac{1}{2})}_{0}$ correspond to the theory
under consideration in flat space. In this section we demonstrate how
they can be found starting from the renormalization group equation.

Let us begin with finding $V^{(0)}_{0}.$ The equation for this
quantity has the form
\beq
\left\{\,\mu\frac{\pa}{\pa\mu}+\be_P
\,\frac{\pa}{\pa P}+
\,\,\ga_\ph \,\ph\, \frac{\pa}{\pa \ph }
\,\right\}\,V^{(0)}_{0}(\ph,P,\mu)\,=\,0. \qquad
\label{RGeffpot-0}
\eeq
The equation (\ref{RGeffpot-0}) is a complicated partial differential
equation with non-constant coefficients. Before solving this equation,
it proves helpful to bring qualitative considerations that simplify the
solution. The parameter $\mu$ can enter to the solution for
$V^{(0)}_{0}$ only logarithmically. Since the argument of the
logarithm must be dimensionless, the dependence on $\mu$ should
be through the parameter $t=\frac{1}{2}\ln\frac{X}{{\mu}^{2}},$
where the quantity $X$ has the mass dimension two. This quantity
can be constructed only from the dimensional parameters of the
classical action, i.e. from $m, M, \ph, g, \tau$ with arbitrary
dimensionless coefficients. In principle, these coefficients should be
fixed with the help of appropriate renormalization conditions for
the effective potential. However, it is natural to assume that the
form of effective potential should be consistent with the form of the
operator ${\cal\hat{H}}$ (\ref{H}) in the scalar sector. It means that
the most natural choice for $X$ is
$X^{(0)}=m^2+g\ph + \frac{1}{2}\la\ph^2$. Thus, we identify
\beq
t^{(0)}\,=\,\frac{1}{2}\ln\frac{m^2+\frac{1}{2}\la\ph^2+g\ph}{\mu^2}
\label{t}
\eeq
for the scalar field contribution to effective potential. The parameters
$M$ and $h$ do not contribute in the scalar sector. Therefore
$V^{(0)}_{0}= V^{(0)}_{0}(t, m^2, g, \tau, \ph)$ and the equation
(\ref{RGeffpot-0}) becomes
\beq
\Big\{ \,\mu\frac{\pa}{\pa\mu}&+&\be^{(0)}_{m^2}
\,\frac{\pa}{\pa m^2}+ \be^{(0)}_{\la}\,\frac{\pa}{\pa \la}
+\be^{(0)}_{g}\,\frac{\pa}{\pa g}
+ \be^{(0)}_{\tau}\,\frac{\pa}{\pa \tau} \,
+ \,\,\ga^{(0)}_\ph \,\ph\, \frac{\pa}{\pa \ph }
\,\Big\}\,V^{(0)}_{0}\,=\,0. \qquad
\label{RGeffpot-0-0}
\eeq
Here the functions
$\be^{(0)}_{\la}$, $\be^{(0)}_{g}$, $\be_{{m}^2}$, $\ga^{(0)}_{\ph}$
and $\be^{(0)}_{\ta}$ are taken at $M=0$ and $h=0$.
For the scalar field contribution $\ga^{(0)}_{\ph}=0$. The next  step
is to express the derivative with respect to $\mu$ through the
derivative with respect to the parameter $t^{(0)}$ defined in (\ref{t}).
As a result, the equation (\ref{RGeffpot-0-0}) looks like
\beq
\Big\{\frac{\pa}{\pa t^{(0)}}&-&\bar{\be}^{(0)}_{m^2}
\,\frac{\pa}{\pa m^2}- \bar{\be}^{(0)}_{\la}\,\frac{\pa}{\pa \la}
-\bar{\be}^{(0)}_{g}\,\frac{\pa}{\pa g}
- \bar{\be}^{(0)}_{\tau}\,\frac{\pa}{\pa \tau}
- \,\,\bar{\ga}^{(0)}_\ph \,\ph\, \frac{\pa}{\pa \ph }
\,\Big\}\,V^{(0)}_{0}\,
=\,0,
\label{RGeffpot-0-0-1}
\eeq
where
\beq
\big(\bar{\be}^{(0)}_{m^2},\,\,
&\bar{\be}^{(0)}_{\la},&\,\,
\bar{\be}^{(0)}_{g},\,\,\bar{\be}^{(0)}_{\tau},\,\,\bar{\ga}^{(0)}_\ph
\big) \,
= \, \frac{1}{1-Q^{(0)}}
\big(
\be^{(0)}_{m^2},\,\,\be^{(0)}_{\la},\,\,\be^{(0)}_{g},\,\,\be^{(0)}_{\tau}
,\,\,\ga^{(0)}_\ph\big)
\label{bar} \qquad
\eeq
and
\beq
Q^{(0)}
= 1-\frac{\pa t^{(0)}}{\pa m^2}
- \frac{\pa t^{(0)}}{\pa \la}
- \frac{\pa t^{(0)}}{\pa \ph}
- \frac{\pa t^{(0)}}{\pa g}.
\label{Q}
\eeq

Solution to the equation (\ref{RGeffpot-0-0-1}) is written as follows
\beq
&V^{(0)}_{0}&(t^{(0)},\,m^2,\,\la,\,g,\,\tau,\,\ph)
= V_{0\,\,cl}(m^2(t^{(0)}),\,\la(t^{(0)}),\,g(t^{(0)}),\,\tau(t^{(0)}),
\,\ph(t^{(0)}),
\label{solution-0} \qquad
\eeq
where
\beq
V_{0\,\,cl}=\frac{1}{2}m^2 \ph^2+\frac{\la}{4!}\ph^4
+\frac{g}{3!}\ph^3+\tau \ph
\label{Vcla}
\eeq
 is the classical potential and
 $m^2(t^{(0)})$, $\la(t^{(0)})$, $g(t^{(0)})$, $\tau(t^{(0)})$ and
 $\ph(t^{(0)})$ are the running parameters $P(t^{(0)})$ and the
 scalar field, satisfying the equations
\beq
\frac{d P(t^{(0)})}{d t^{(0)}}
&=&
\bar{\be}^{(0)}_{P}(t^{(0)}),
\nn
\\
\frac{d \ph(t^{(0)})}{d t^{(0)}}
&=&
\bar{\ga}_{\ph}(t^{(0)})
\label{P}
\eeq
with the  initial conditions
\beq
P(t)_{\vert{t=0}} &=& P.
\label{initial}
\eeq
As before, here $P={m^2,\,\la,\,g,\,\tau}$. Since we work in the
one-loop approximation, all quantum corrections are linear in
$\hbar$, hence we can set the quantity $Q$ (\ref{Q}) equal to zero
in the expressions for beta- and gamma-functions. Then the solutions
of the equations (\ref{P}) can be easily found
\beq
P(t^{(0)}) &=&P+\be^{(0)}_{P}t^{(0)},
\nn
\\
\ph(t^{(0)}) &=& \ph + \ga^{(0)}_{\ph}t^{(0)}=\ph,
\label{running}
\eeq
where we took into account that $\ga^{(0)}=0.$ The relations
(\ref{running}) together with the explicit forms  for the functions
$\be^{(0)}_{P}$ represent the solution for the effective potential
$V^{(0)}_{0}$.

The analysis of the $V^{(0)}_{1}$ can be done in a similar way,
so we skip the details.  The result has the following form
\beq
V^{(0)}_{1}=V_{1\,cl}(P_{1}(t^{(0)}),\,\ph(t^{(0)}))
\label{V-0-1}
\eeq
with $V_{1\,cl}R=-\frac{1}{2}\xi R \ph^2+f R \ph$. The quantities
$P_{1}(t^{(0)})=P_{1}+\be^{(0)}_{P_{1}}t^{(0)} $ and $\be^{(0)}_{P_{1}}=(\be^{(0)}_{\xi},\,\be^{(0)}_{f})$. These
relations together with (\ref{V-0-1}) are final solutions for
curvature dependent contribution to effective potential from
quantum scalar field.

We now turn to finding the quantum contribution $\bar{V}^{(\frac{1}{2})}_{0}+R\bar{V}^{(\frac{1}{2})}_{1}$ to effective
potential from quantum spinor field. In this case we begin with equation
(\ref{RGeffpot}) for $V^{(\frac{1}{2})}_{0}$ and $V^{(\frac12)}_1$
separately taking into account that consistence with form of the operator
$\hat{H}$ in fermionic sector (\ref{hatH}) motivates a natural choice for
dimensionless parameter containing the logarithm of $\mu$  in the form
\beq
t^{(\frac{1}{2})}=\frac{1}{2}\ln \frac{(M+h\ph)^2}{\mu^2}.
\label{t-1}
\eeq
All other considerations are analogous to one for $V^{(0)}_{0}$ and
$V^{(0)}_{1}$. Thus,  we present only the final results for quantum
corrections,
\beq
\bar{V}^{(\frac{1}{2})}_{1}
\,=\, V_{0\,\,cl}(P^{(\frac{1}{2})}(t^{(\frac{1}{2})}),
\,\ph(t^{(\frac{1}{2})})),
\label{barV-0}
\eeq
and
\beq
R\bar{V}^{(\frac{1}{2})}_{1}
\,=\, RV_{1\,cl}(P^{(\frac{1}{2})}_{1}(t^{(\frac{1}{2})}),\,
\ph(t^{(\frac{1}{2})})),
\label{barV-1}
\eeq
where the running parameters and field have the form
\beq
P^{(\frac{1}{2})}(t^{(\frac{1}{2})})
&=&
\be^{(\frac{1}{2})}_{P}t^{(\frac{1}{2})},
\nn
\\
\ph(t^{(\frac{1}{2})})
&=&
\ga^{(\frac{1}{2})}_{\ph}t^{(\frac{1}{2})}.
\label{running-1}
\eeq
Here we have solved the equations for running parameters and field
with zero initial conditions since the classical contribution to
effective potential has been already found when we calculated
$\,V^{(0)}_{0}$ and $V^{(0)}_{1}$.
The functions $\be^{(\frac{1}{2})}_{P}$ and
$\ga^{(\frac{1}{2})}_{\ph}$ in the relations (\ref{running-1}) are
the $\be_{P}$ and $\ga_{\ph}$ at non-zero $M$ and $h$ but with
zero parameters $m^2,\,\la,\,g,\,\tau,\,\xi.$ The relations (\ref{barV-0})
and (\ref{barV-1}) define the final contribution to effective potential
from quantum spinor field.

%%%   red
%%%  {\LARGE \textbullet}
Thus, we are in a position to present an explicit expression
for the Minimal Subtraction scheme - based effective potential,
\beq
V_{eff}&=& - \frac12 \, m^{2} \ph^2
-  \frac12 \, \xi R \ph^2
+ \frac{\la}{4!}\,\ph^4
+  \frac{g}{3!} \ph^3
+  \tau \ph
+  fR \ph
\nn
\\
&-&
\frac{1}{2(4\pi)^2} \Big\{
\Big[12 N M^2 h^2t^{(\frac{1}{2})}
\,+\,\frac{\la m^2 - g^2}2\,t^{(0)}  + C_1 \Big] \ph^2
\nn
\\
&+& \big[ g m^2t^{(0)} + 8 N h M^3t^{(\frac{1}{2})} + C_5\big] \ph
+
\Big[ \frac{\la}2\, \Big(\xi - \frac{1}{6}\Big) t^{(0)}
\,-\,\frac{Nh^2}{3}\,  t^{(\frac{1}{2})}  + C_2\Big] R\ph^2
\nn
\\
&-& \frac{1}{4!}\,\big[3 \la^2 t^{(0)}  - 48 N h^4 t^{(\frac{1}{2})} + C_4
\big]\ph^4
\nn
\\
&-&
\frac{1}{3!}\,
\Big[\frac{3\la g}{2}\,\,t^{(0)}
- 3N \big( g h^2 - 4 M h^3\big) t^{(\frac{1}{2})} + C_3\Big] \ph^3
\nn
\\
&+&
\Big[ \frac{2N M h}{3} \, t^{(\frac{1}{2})}
- g \Big(\xi -\frac{1}{6}\Big)t^{(0)} + C_6\Big] R\ph \Big\},
\label{VeffMS}
\eeq
where $\,t^{(0)}\,$ and $\,t^{(\frac{1}{2})}\,$ were identified
in $(\ref{t})$ and (\ref{t-1}). The constants $C_{1\,...\,6}$ can be
found from the initial renormalization conditions. For instance, the two
well-known values which correspond to the standard choices in the
massless scalar case are $C_4=-\frac{25}{6}$ obtained in
\cite{ColeWein} and $C_2=-3$ obtained in \cite{BuchOd84}
(see \cite{book} for a more pedagogical derivation). Since this
calculation of the values of  $C_{1\,...\,6}$ for the massive theory
is rather cumbersome and there are no immediate applications, we
skip it. Let us stress that the quantum corrections in
Eq.~(\ref{VeffMS}) have qualitatively new terms with odd powers
of a sterile scalar field, multiplied by the two kinds of logs. The
final definition of the corresponding renormalization constants
$C_{3,5,6}$ requires independent measurements and can be achieved
only within an appropriate experimental or observational framework.
Some of the possible observables related to the odd terms will be
discussed in Sec.~\ref{sectSSB}.
%%%  red
%% \textcolor{red}{\LARGE \textbullet}

%%%%%%%%%%%%%%%%%%%%%%%%%%%%%%%%%%%%%%%%
\section{Direct calculation of effective potential}
\label{sect4}

The results of the previous sections have shown the importance of
the terms which are odd in the scalar field. This is something we
learned from the divergences derived in the framework of the
Scwinger-DeWitt method. Due to the importance of this quantum
calculation, it looks reasonable to control its output by qualitatively
different method. This is done in the present section by deriving
effective potential in the ${\cal O}(R)$-approximation using the
normal coordinates and local momentum representation, in a way
similar to what was done recently in \cite{CorPot}, where one can
find many relevant technical details and further references.

%%%%%%%%%%%%%%%%%%%%%%%%%%%%
%%%%%%%%%%%%%%%%%%%%%%%%%%%%
\subsection{Riemann normal coordinates and scalar contribution}

These coordinates are related to the geodesic lines which link a
fiducial point $P'(x^{\mu'})$ with another point with the
coordinates $\,x^{\mu'}=x^\mu + y^\mu$. In order to use the local
momentum representation we assume that
$g_{\mu \nu} (P') = \eta_{\mu\nu}$. In the vicinity of this point,
in the linear in curvature approximation, we have
\beq
g_{\alpha \beta} (x) = \eta_{\alpha \beta} (x')
- \frac{1}{3} R_{\alpha \mu \beta \nu} (x') y^\mu y^\nu
\,+\, \cdots\,.
\eeq
Then, the bilinear operator in the scalar sector can be written as
\beq
-\,\hat{H}
&=&
\frac{1}{\sqrt{-g}}
\frac{\delta ^2 S_{scalar}}{\delta \ph (x) \delta \ph (x')}
\,=\,
\Box + V'',
\n{2}
\eeq
where $V''$ is the second derivative of the classical potential
\beq
V(\ph) =
\frac{1}{2} m^2 \ph^2
+ \frac{\lambda}{4!} \ph ^4
+ \frac{g}{3!} \ph ^3 + \tau \ph + fR.
\label{V}
\eeq
Further calculation in this subsection will essentially repeat the
one of \cite{CorPot}, but with another potential (\ref{V}). We
include this short review part for making all the presentation more
consistent.

We can expand (\ref{2}) in the Riemann normal coordinates as
\beq
-\hat{H} =
\eta ^{\mu \nu} \partial_\mu \partial_\nu
+ \frac{1}{3} R^{\mu}{}_\alpha{}^\nu{}_\beta y^{\alpha} y^{\beta}
\pa_\mu \pa_\nu
- \frac{2}{3}R^{\alpha}{}_\beta y^\beta \partial_\alpha + m^2 - \xi R + V''
\,+\, \cdots\,.
\eeq
where $\cdots$ denotes high order terms in curvature.

The main advantage of the local momentum representation is that the
calculation can be performed in flat space-time and the result can be
always presented in a covariant way. For instance, the equation for the
propagator of  a real scalar field has the form
\beq
\hat{H}G(x,x') = - \delta ^c (x,x'),
\eeq
where $\de^c(x,x') = g^{-\frac{1}{4}}(x') \delta (x,x')g^{-\frac{1}{4}}(x)$
is a covariant Dirac delta function.

Since we are going to make calculations around the flat metric, it is
most useful to work with the modified propagator $\bar{G}(x,x')$
where
\beq
\hat{H}\bar{G}(x,x') = - \delta (x,x')
\nn
\eeq
The explicit form of $\bar{G}(x,x')$  is known
\cite{BunchParker79,ParkerToms} for the free case when $V''=m^2$
and for the even potential at constant $\ph$ \cite{CorPot}. As far
as it is sufficient to regard $V'' = const$ for the derivation of
effective potential, we can replace $m^2$ by $\tilde{m}^2 = V''$
and obtain, in first order of curvature expansion, the following
expression:
\beq
\bar{G}(y)
=
\int \frac{d^4 k}{(2\pi)^4}e^{ikx}\Big[\frac{1}{k^2 - \tilde{m}^2}
- \Big(\xi - \frac{1}{6} \Big) \frac{R}{(k^2 - \tilde{m}^2)^2}\Big]
\,+\, \cdots\,.
\eeq
We can expand $\hat{H}$ and $\bar{G}(x,x')$ up to the first power
of  scalar curvature as
\beq
\hat{H} &=&
\hat{H}_0 + \hat{H}_1 R + O(R_{\dots}^2)\,,
\nn
\\
\bar{G}&=& \bar{G}_0 + \bar{G}_1 R + O(R_{\dots}^2)\,.
\nn
\eeq
Starting from this point the ${\cal O}(R_{\dots}^2)$ terms will not be
mentioned.

As far as $\,\Tr \ln \hat{H} = -\Tr \ln G(x,x')$, we get
\beq
-\frac{1}{2} \Tr \ln \bar{G}(x,x')
&=&
\frac{1}{2} \Tr \ln (\hat{H}_0 + \hat{H}_1 R)
\nn
\\
&=&
\frac{1}{2} \Tr \ln \hat{H}_0
+ \frac{1}{2}\Tr(\bar{G}_0\hat{H}_1 R).
\label{expa} \quad
\eeq

Consider first the effective potential in flat space. The first term
in the {\it r.h.s} of (\ref{expa}) includes the flat space
contribution $ \bar{V}_0(\ph)$, which can be defined as
\beq
\bar{V}_0(\ph) =
\frac{1}{2}\Tr \ln S_2 (\ph)
- \frac{1}{2}\Tr \ln S_2(\ph = 0).
\n{a}
\eeq
Here $S_2$ is the bilinear form of the classical scalar field action,
\beq
S_2 (\ph) \,=\,
\frac12
\int d^4 x \Big\{\ph\eta^{\mu\nu} \partial_\mu \partial_\nu \ph + V''
\Big\}.
\eeq
As a result we have
\beq
\bar{V}_0 (\ph) =
\frac{1}{2} \Tr \ln \Big\{ \frac{\Box + V''}{\Box +m^2}
\Big\}.
\eeq

Unlike the previous sections, all the subsequent calculations will
be performed in the cut-off regularization what helps to simplify
the calculations and make them more explicit. The transition to
dimensional regularization can be easily done, of course. By
introducing the four-dimensional Euclidean momentum cut-off
$\Omega$ and integrating over angular coordinates, we arrive at
\cite{CorPot}
\beq
\bar{V}_0 (\ph)
\,=\,
\frac{1}{2(4\pi)^2}
\int_0 ^\Omega k^2 dk^2 \ln \Big( \frac{k^2 + V''}{k^2 + m^2}\Big).
\eeq
Taking the last integral, after some algebra we obtain
\beq
&&\bar{V}_0 (\ph, \eta_ {\mu \nu}) =
\bar{V}_0 ^{div} + \bar{V}_0 ^{fin},
\nn
\\
&&V_0 ^{div} =
\frac{1}{2(4\pi)^2} \Big\{\Omega ^2 V''
- \frac{1}{2}(V'')^2 \ln \frac{\Omega^2}{m^2} \Big\},
\nn
\\
&&\bar{V}_0 ^{fin} =
\frac{1}{2(4\pi)^2}
\Big\{ \frac{1}{2}(V'')^2 \ln\Big(1+ \frac{V''}{m^2}\Big)
- \frac{1}{4}(V'')^2 \Big\}. \qquad
\eeq
In order to cancel divergences, we follow the minimal substraction
scheme and introduce an appropriated counterterm in the form
\beq
\Delta V_0
=
\frac{1}{2(4\pi)^2} \Big\{
-\Omega ^2 V''
+ \frac{1}{2}(V'')^2\ln \frac{\Omega^2}{\mu ^2}
+ \frac{1}{4} (V'')^2\Big\},
\eeq
where $\mu$ is the dimensional renormalization parameter. In this way,
the quadratic and logarithmic divergences are eliminated and the
renormalized effective potential can be written as
\beq
V_{eff} ^{ren} (\eta_{\mu \nu}, \ph) =
V + \bar{V}_0 + \Delta V_0
=
%\,=\,
V + \frac{(V'')^2}{64 \pi ^2}\,
\ln \Big(\frac{V''}{\mu ^2}\Big).
\label{V0scal}
\eeq

Let us consider the linear in curvature corrections. The first order
contribution is due to the second term in the \textit{r.h.s.} of
Eq.~(\ref{a}). This term can be easily presented in the form
\beq
\frac{1}{2} \Tr (\bar{G}_0\hat{H}_1 R)
%% &=&
%% - \int d^4x V_1 R
%% = \frac{1}{2}\Tr[\bar{G}^{-1} _0 (x'',x')\bar{G}_1 (x',x)]R
%% \nn \\
&=& \frac{1}{2} \int d^4 x
\int d^4 x' [\bar{G}^{-1} _0 (x,x')\bar{G}_1 (x',x)]R
\nn
\\
&=& \frac{1}{2} \int d^4 x \int d^4 x' R
\int \frac{d^4k}{(2\pi)^4} e^{ik(x-x')}
\times
\int \frac{d^4p}{(2\pi)^4} e^{ip(x'-x)}
\bar{G}^{-1}_0 (k) \bar{G}_1(p)
\nn
\\
&=&
\frac{1}{2} \int d^4 x \, R \int \frac{d^4k}{(2\pi)^4}
\bar{G}_0 ^{-1} (k) \bar{G}_1 (k)
\eeq
and hence
\beq
&&\frac{1}{2} \Tr (\bar{G}_0\hat{H}_1 R)=
- \frac{1}{2(4\pi)^2}\Big( \xi - \frac{1}{6} \Big)
\int d^4 x \, R
\int_0^{\Omega}  \frac{k^2dk^2}{k^2 + \tilde{m}^2}.
\eeq
After taking the last integral, the final result reads
\beq
\bar{V}(\ph , g_{\mu \nu}) = \bar{V}_0 + \bar{V}_1 R,
\eeq
where $\bar{V}_0$ is given by Eq.~(\ref{V0scal})
and $\bar{V}_1 = \bar{V}_1 ^{fin} + \bar{V}_1 ^{div}$, where
\beq
\bar{V}_1 ^{div} &=&
\frac{1}{2(\pi)^4}
\Big( \xi - \frac{1}{6} \Big)
\Big[ - \Omega ^2 + (V'')\ln \frac{\Omega ^2}{m^2} \Big]\,,
\nn
\\
\bar{V}_1 ^{fin} &=&
\frac{1}{2(\pi)^4}
\Big( \xi - \frac{1}{6} \Big)
\Big[(V'') \ln \Big( \frac{V''}{m^2}\Big) \Big]\,.
\eeq
Similar to the flat space case, the potential must be modified by
adding a counterterm,
\beq
\Delta\bar{V}_1 =
\frac{1}{2(\pi)^4} \Big( \xi - \frac{1}{6} \Big)
\Big[ \Omega^2 - (V'') \ln \frac{\Omega^2}{\mu^2} \Big]\,.
\eeq
Thus, the renormalized expression is
\beq
V_{eff,1} ^{ren} (g_{\mu \nu} , \ph)
&=&
\frac{1}{2}(m^2\,-\,\xi R )\ph ^2
- \frac{1}{2(\pi)^4}
\Big(\xi - \frac{1}{6} \Big) (V'') \ln
\Big(\frac{V''}{\mu ^2} \Big). \quad
\n{5}
\eeq
The full renormalized effective potential for the scalar sector
of (\ref{1}) is the sum of expressions  (\ref{V0scal}) and (\ref{5}),
\beq
V_{eff} ^{ren} (g_{\mu \nu},\ph)
&=&
\rho _{\Lambda} + V + \frac{\hbar}{2(4\pi)^2} \Big[ \frac{1}{2} (V'')^2
- \Big( \xi - \frac{1}{6} \Big)R (V'') \Big]
\ln \Big( \frac{V''}{\mu ^2}\Big),
\label{VrenScal}
\eeq
where we restored the first power of the loop parameter $\hbar$
and added the cosmological constant term, $\rho_\La$.

%%%%%%%%%%%%%%%%%%
%%%%%%%%%%%%%%%%%%
%%%%%%%%%%%%%%%%%%
\subsection{Fermion contributions and overall expression}

Let us consider now the fermion contribution to the
effective potential of the sterile scalar. In the case of potential
the background field $\ph$ can be treated as a constant, hence
we denote $\tilde{M} = M + h\ph$. Taking the Grassmann parity
of the quantum field into account, in the Euclidean notations we get
\beq
\Gamma_f ^{(1)} [\ph , g_{\mu \nu}]
\,=\,
- \,\Tr \ln \hat{H}_f,
\label{trfer}
\eeq
where $\hat{H}_f = i(\ga^\mu \na_\mu + i \tilde{M})\de^{ij}$.
As usual, we consider \cite{GBP}
\beq
\Tr \ln \hat{H}_f =
\frac{1}{2} \Tr \ln (\hat{H}_f \hat{H}_f ^*),
\eeq
with  $\hat{H}_f ^* = i(\gamma ^\mu \nabla _\mu - i \tilde{M})\delta _{jk}$.
After some algebra this gives
\beq
\Tr \ln \hat{H}_f =
\frac{1}{2} \Tr \ln
\Big(-\Box + \frac{1}{4}R - \tilde{M}^2 \Big)\delta_{k} ^{i}
\eeq
The fermion propagator is defined from the relation
\beq
(\hat{H}_f \hat{H}_f ^*) \mathcal{G}(x,x') = - \delta^c (x,x').
\eeq
Following the same scheme which was used in the scalar case,
one can define modified propagator
\beq
\mathcal{\bar{G}}(x,x')
\,=\, \mathcal{\bar{G}}(x,x') g^{-\frac{1}{4}}(x),
\eeq
which satisfies the equation
\beq
(\hat{H}_f \hat{H}_f ^*) \mathcal{\bar{G}}(x,x')
= - \delta (x,x').
\eeq
From the paper by Bunch and Parker \cite{BunchParker79}
we learn that $\mathcal{\bar{G}}(x,x')=\mathcal{\bar{G}}(y)$
is defined as
\beq
\mathcal{\bar{G}}(y) =
\int \frac{d^4k}{(2\pi)^4} e^{iky}
\Big[ 1 - \frac{1}{12}R \frac{\partial}{\partial \tilde{M}^2} \Big]
(k^2 + \tilde{M}^2)^{-1} \hat{1} \quad
\eeq
in the first order in curvature.
Thus,
\beq
-\frac{1}{2}\Tr \ln \mathcal{G}(x,x')
= \frac{1}{2}\Tr \ln \hat{H}_f
= \frac{1}{4} \Tr \ln (\hat{H}_f \hat{H}_f ^*). \quad
\eeq
Using the same considerations as for the scalar field, we find that
\beq
-\frac{1}{2}\Tr \ln \mathcal{G}(x,x') =
\frac{1}{4}\Tr \ln (\hat{H}_f \hat{H}_f ^*)_0
+ \frac{1}{4} \Tr \bar{\mathcal{G}}_0 (\hat{H}_f \hat{H}_f ^*)_1 R \n{6}
\eeq
The first term in \textit{r.h.s.} correspond to the flat space case
and the second one is the first order in curvature contribution. We
will first perform the calculation in the flat space, when
\beq
\frac14\Tr \ln (\hat{H}_f \hat{H}_f ^*)_0
\,=\,\frac14\, \sTr \ln (-\eta ^{\mu \nu} \pa_\mu \pa_\nu
- \tilde{M}^2) \,\delta ^i _k .
\nn
\eeq
In the momentum representation this gives
\beq
\frac{1}{4}\Tr \ln (\hat{H}_f \hat{H}_f ^*)_0
\,=\,
(-2N) \int_0 ^\Omega \frac{d k^2}{(4\pi)^2}k^2
\ln \Big(\frac{1}{k^2 + \tilde{M}^2}\Big),
\nn
\eeq
which provides the flat space part of the one-loop
effective potential,
\beq
V_0 ^{div} (fer)
\,=\,
- \frac{2N}{(4\pi)^2}
\Big\{ \frac{1}{2} \ln \Big(\frac{\Omega^2}{\tilde{M} ^2}\Big)
\tilde{M^4} + \frac{1}{4} \Omega^2 \Big\}.
\eeq
In order to renormalize this result, we introduce a counterterm of the form
\beq
\Delta V_0 =
\frac{2N}{(4\pi)^2} \Big\{ \frac{1}{2} \ln
\Big(\frac{\Omega^2}{\mu ^2}\Big)\tilde{M^4}
+ \frac{1}{4} \Omega^2 \Big\}
\eeq
Thus,
\beq
 V_0 ^{ren} (fer) =
\frac{N}{(4\pi)^2}\ln \Big(\frac{\tilde{M} ^2}{\mu ^2}\Big)\tilde{M^4}
\eeq
The contribution in first order of curvature is
\beq
\frac{1}{4} \Tr \bar{\mathcal{G}}_0 (\hat{H}_f \hat{H}_f ^*)_1 R
=
(-2N) \int d^4 x R
\times
\int \frac{d^4k}{(2\pi)^4} \bar{\mathcal{G}}_0 ^{-1} (k)
\bar{\mathcal{G}}_1 (k).
\eeq
which can be written in momentum space as
\beq
\frac{1}{4} \Tr \bar{\mathcal{G}}_0 (\hat{H}_f \hat{H}_f ^*)_1 R \nn
=
- \frac{N}{6(4\pi)^2} R \int_0 ^\Omega dk^2 \frac{k^2}{k^2
+ \tilde{M}^2}.
\nn
\eeq
Therefore,
\beq
V_1 ^{div}(fer) R
\,=\,
\frac{N}{6(4\pi)^2}R
\Big\{\tilde{M}^2 \ln \frac{\Omega^2}{\tilde{M}^2}
+ \Omega ^2 \Big\},
\label{VdivFer}
\eeq
while there is no remnant finite part in this case. The divergences can
be eliminated by adding a counterterm
\beq
\Delta V_1
\,=\,
- \frac{N}{6(4\pi)^2}\Big\{\tilde{M}^2
\ln \frac{\Omega^2}{\mu^2} + \Omega ^2 \Big\}.
\label{DeltaVFer}
\eeq
Finally, the first order in curvature part of the renormalized
effective potential has the form
\beq
V_{ren} (fer)
\,=\,
- \frac{N}{(4\pi)^2}
\Big\{ \tilde{M}^4 - \frac{1}{6} R \tilde{M}^2 \Big\}
\ln \Big(\frac{\tilde{M}^2}{\mu ^2}\Big).
\label{VrenFerm}
\eeq

Summing up the scalar (\ref{VrenScal}) and fermion (\ref{VrenFerm})
contributions, we arrive at the general expression for the effective
potential of our model, which includes a single real sterile scalar
and $N$ copies of massive fermion fields,
\beq
V_{eff} ^{ren} (g_{\mu \nu} , \ph)
&=&
\rho _{\Lambda} + \frac{1}{2} (m^2 - \xi R) \ph ^2 + V
+ \frac{\hbar}{2(4\pi)^2}\Big\{ 
- 2N (M+ h\ph)^4\,\ln \Big[\frac{(M+ h\ph)^2}{\mu^2}\Big]
\nn
\\
&+& \Big[ \frac{1}{2} (V''+ m^2)^2
- \Big( \xi - \frac{1}{6} \Big) R (V'' + m^2) \Big]
\ln \Big( \frac{V'' + m^2}{\mu ^2} \Big)
\nn
\\
&+&
\frac{N}{3}R (M + h\ph)^2 \ln \Big[\frac{(M+ h\ph)^2}{\mu^2}\Big]
\Big\},
\label{VeffTot}
\eeq
where the interacting and odd terms of the classical potential $V$
(remember that we separated the term with scalar mass for the sake
of convenience) are  defined in Eq.~(\ref{V}).
%%%  red
%% {\LARGE \textbullet}
It is easy to verify the perfect correspondence with the expression
(\ref{VeffMS}) derived from the Minimal Subtraction - based
renormalization group with the scale identifications (\ref{t}) and
(\ref{t-1}).
%%%  red
%% \textcolor{red}{\LARGE \textbullet}

The effective potential (\ref{VeffTot}) depends on an arbitrary
parameter $\mu$. To fix the value of this parameter, we should
imposed the renormalization conditions in a usual way.

%%%%%%%%%%%%%%%%%%%%%%%%%%%%%%%%%%
%%%%%%%%%%%%%%%%%%%%%%%%%%%%%%%%%%
%%%%%%%%%%%%%%%%%%%%%%%%%%%%%%%%%%
\section{Induced action of gravity with odd terms}
\label{sectSSB}

In this section we discuss some interesting aspects of the model
under consideration related to symmetry breaking, induced action
of gravity and its possible physical manifestations.

In the scalar theory in flat space without odd terms in the scalar
sector, the classical potential $U=-\frac{1}{2}m^2\ph^2+V$ with
``untrue'' sign at $m^2$ has a constant minimum position, which
corresponds to the spontaneous symmetry breaking (SSB) of
discrete symmetry. In curved space such a constant position of
the minimum and corresponding vacuum state are impossible
\cite{sponta}. The most immediate reason is that the classical
potential contains the non-minimal $\,\xi R(x)\ph^2\,$ term and
constant solution for the scalar field is impossible in the general
case of an arbitrary metric. Such a general analysis is beyond the
scope of this paper, the discussion of related issues can be found
e.g. in Ref.~\cite{Tmn-ABL}.

Things get even more complicated in the case of the sterile scalar,
since the odd powers of the scalar in the potential make the
discrete symmetry impossible and hence there is no much sense to
speak about its breaking. However, let us assume that odd terms
and also the more traditional non-minimal term $\,\xi R\ph^2\,$
are small and treat them as small perturbations. In this case we
have a SSB in the zero-order approximation, and this is the
terminology which we shall adapt in what follows.

Let us consider the equation of motion for vacuum expectation
value (VEV) of the scalar field.
\beq
\frac{\delta S}{\delta \ph}\bigg\vert _{\ph = v} = 0,
\label{eq: vev}
\eeq
where $v$ is supposed to be constant and in many cases it is so.
However, as we have just mentioned, the solution of this problem
in curved
space is quite nontrivial \cite{sponta}, because in general $R$ is not
constant and $\xi \neq 0$. As a result,  the equation for the VEV
\beq
- \Box v + m^2 v + \xi R v
- \frac{1}{6} \lambda v^3  - \frac{1}{2} g v^2- \tau - fR = 0
\label{eq: vev2} \quad
\eeq
can not be solved with constant $v$ even neglecting the kinetic
term. Following \cite{sponta} we can expand the solution into
power series in the curvature tensor or $\xi$, such that
\beq
v(x) = v_0 + v_1(x) + v_2(x) + \dots \,\,.
\eeq
In the solution of the problem of VEV we shall treat both $\xi R$
and odd terms as small perturbations.

In the zero order we have
\begin{eqnarray}
v_0 ^2 = \frac{6m^2}{\lambda}\,.
\end{eqnarray}
Now, in order to solve Eq.~(\ref{eq: vev2}) in the first order,
consider the following approximations:
\beq
\vert g \vert  \ll  v_0,
\quad \vert \tau \vert \ll {v_0}^{3},
\quad \vert \xi R \vert \ll {v_0}^{2},
\quad \vert f \vert \ll v_0.
\quad
\eeq

After solving the perturbative problem independently for $\xi R$
and odd terms, and summing up the results we obtain
\beq
v(x)
&=&
v_0 - \frac{\tau}{\mu_0^{2} + g v_0}
\,+\,\frac{\xi v_0 - f}{\Box + \mu_0^2}\, R,
\n{2a}
\eeq
where
\beq
\mu_0 ^2 = \frac{\lambda}{3} v_0 ^2.
\n{3}
\eeq

To obtain the induced low-energy action we substitute the solution
(\ref{2a}) into the action (\ref{1}). The result has the form
\beq
S_{ind} &=&
\int d^4 x \sqrt{-g}
\Big\{
- \rho_{\Lambda} ^{ind} - \frac{1}{16 \pi G_{ind}} R
+ \Big(\tau f - \tau \xi v_0 \Big) \frac{1}{\Box + \mu_0 ^2} R
\nn
\\
&+&
\Big( f\xi v_0 - \xi^2 v_0 ^2
- frac{1}{3} g \xi v_0 ^3
+ \frac{1}{3} g f v_0 ^2 - f \xi v_0 + f^2 \Big)
R \frac{1}{\Box + \mu_0 ^2} R  \Big\}.
\n{4} \qquad \quad
\eeq
In the expression (\ref{4}) the cosmological constant and the
inverse Newton constant are defined by the expressions
\beq
\rho_{\Lambda} ^{ind} &=&
- \frac{\lambda}{24} v_0 ^4
- \frac{4}{3} \,\frac{\tau \lambda v_0 ^3}{\mu_0 ^2 + gv_0}
+ \frac{1}{6}\, g v_0 ^3
+ \tau v_0 - \frac{\tau ^2}{\mu_0 ^2 + gv_0},
\\
\frac{1}{16 \pi  G_{ind}}
&=&
\frac{\tau v_0 \xi}{\mu_0 ^2 + g v_0}
- \frac{\xi v_0}{2} + f v_0 - \frac{\tau f }{\mu_0 ^2 + g v_0}.
\label{ind} \qquad
\eeq
As expected, these formulas
show the small contributions of the odd parameters of the sterile
scalar $\tau$, $f$ and $g$. Since these parameters are certainly small,
the change in these induced quantities is irrelevant compared to the
quantities induced, e.g., in the electroweak phase transition, where
$v_0$ is the vacuum expectation value of the Higgs field and the
mentioned induced quantities may be much larger that, for instance,
for the quintessence field.

At the same time  the induced action  (\ref{4}) have two other details
which may be in fact more significant. First of all, in the case of
quintessence the mass scale should be very small, and hence the
VeV value $v_0$ should be small too. Then the non-local terms in
the second line of Eq.~(\ref{4}) may become phenomenologically
relevant. It would be interesting to explore the phenomenological
limits on the odd parameters,
starting from $v_0$ from the experimental data on Newton law and
observational data on the bending of light, and see whether these limits
can produce some restrictions on the quintessence potential. This
investigation is beyond the scope of the present work, since it is
devoted to the quantum aspects of a sterile scalar, however even
the possibility looks attractive.

Another aspect concerns the term  $(\Box + \mu_0^2)^{-1} R$
in the first line of Eq.~(\ref{4}). According to the recent discussion
in Ref.~\cite{Omar-FF} this term can be relevant in cosmology,
in the periods when the inverse of the size of the horizon may be
comparable with the cosmic scale. Up to some extent, the corresponding
effect can lead to the change of observational predictions in both
inflationary and late cosmology epochs.

%% red
%% {\LARGE \textbullet}
Let us now consider an application of odd terms in the scalar action
(\ref{classact}) to inflation. Since the present-day reference theory
for inflation is the one of Starobinsky \cite{star}, the simplest
approach is to make a mapping to the $R+F(R)$ and take care about
the effect of the odd terms. As usual, we consider that the
non-minimal term and the $\la\ph^4$-terms in the classical potential
(\ref{Vcla}) are dominating, while the mass $m^2\ph^2$-term,
classical odd terms and the leading quantum corrections in the
effective potential (\ref{VeffMS}) are small and can be regarded
as perturbations. As a first approximations and for the sake of
simplicity, the kinetic $(\pa\ph)^2$-term can be completely omitted,
assuming that $R$ is almost constant in the inflationary period.
Let us stress that the detailed analysis of inflation in the model
under consideration is beyond the scope of the present work, we
are just trying to sketch the main features of the inflationary model
with odd terms, and leave the rest of the work for the future.

The derivation of the induced action of gravity in this approximation
differs from what we have done above. The equation for the scalar
field is
\beq
\xi R\ph - \frac16\,\la \ph^3 - m^2\ph - \frac12\,g \ph^2 -
\tau - fR\,=\,0.
\mbox{\quad}
\label{inf1}
\eeq
In the zero-order approximation we get
\beq
\xi R\ph_0 - \frac16\,\la \ph_0^3 \,=\,0
\quad
\Longrightarrow
\quad
\ph_0^2 = \frac{4\xi R}{\la}.
\mbox{\quad}
\label{inf2}
\eeq
Let us remark that since during inflation $R$ is negative and we
need a positive $\la$ for the stability of the potential, this
solution implies $\xi < 0$, such that the product $\xi R > 0$.
Substituting (\ref{inf2}) back into the
action, in the leading order there is an induced term
$\,\frac{3\xi^2}{2\la}R^2$. According to the classical estimate
of \cite{star83}, this means that the ratio $\,\frac{3\xi^2}{2\la}\,$
should be close to $5 \times 10^8$ to provide a successful model
of inflation.

In the next approximation we consider $\ph = \ph_0 +  \ph_1$, and
arrive at the linear equation for $\ph_1$,
\beq
\Big( \xi R - \frac{\la}{2}\,\ph_0^2\Big)\ph_1
\,=\, m^2\ph_0 + \frac{g}{2}\,\ph_0^2 + \tau + fR,
\mbox{\quad}
\label{inf3}
\eeq
which solves in the form
\beq
\ph_1
\,=\, - \,m^2\sqrt{\frac{3}{2\la \xi R}}
\,\,-\,\frac{3g}{2\la}\,-\,\frac{\tau}{2\xi R}\,-\,\frac{f}{2\xi}\,.
\mbox{\quad}
\label{inf4}
\eeq
After placing the sum $\ph = \ph_0 +  \ph_1$ into the action,
in the leading order we get the induced Lagrangian of gravity that
corresponds to the given approximation,
\beq
L_{ind}
&=&- \, \frac{3m^2\xi}{\la}\,R
\,+\,\frac{3\xi^2}{2\la}\,R^2
+ \sqrt{\frac{6 \xi R}{\la}}
\Big[\tau  \,-\,  \Big(f + \frac{g\xi}{\la}\Big)R\Big].
% \mbox{\quad}
\label{inf5}
\eeq
The first term in the this expression is a small irrelevant
addition to the classical Einstein-Hilbert term $\sim M_P^2R$.
The second term in the first line is the leading $R^2$-term which
was mentioned above, it represents the main element of the
Starobinsky inflation \cite{star,star83}. According to our
approximation, the terms in the second line of (\ref{inf5}) represent
a small addition to the main  $R^2$-term. The analysis of the
effect of these terms on the inflationary observables (mainly the
spectral index $n_s$ of the primordial curvature perturbations
and the tensor-to-scalar ratio $r$) can be done along the way
of the previous works in this direction \cite{Huang,AR-RbR}.

The comparison to the polynomial model of perturbations in
\cite{Huang} is especially instructive. Let us stress that the odd
terms in the {\it classical} potential of the sterile scalar field are
necessary to provide a {\it quantum consistency} of the theory
of a sterile scalar coupled to fermions. Our consideration shows
that this implies the induced gravitational action to include the
non-polynomial terms (\ref{inf5}).  As we mentioned above,
the detailed analysis of these terms is beyond the scope of the
present work. However, we can use the renormalization group
equations (\ref{betas}) for the parameters of the action,
together with the estimate \cite{star83} (see also recent paper
\cite{StabInstab}) for the ratio
$\,\frac{3\xi^2}{2\la}\propto 5 \times 10^8$, to
evaluate the magnitudes of the induced non-polynomial terms.

Assuming that $\la \approx 1$, we arrive at the estimate
$|\xi| \approx 4 \times 10^4$, similar to the Higgs inflation
\cite{Shaposh}. Next, according to the equations  (\ref{betas}),
the lower bound for the absolute values of the odd parameters
is defined by the mass of the heaviest fermion that couples to
the sterile scalar, multiplied by the corresponding Yukawa
coupling. Assuming that this coupling is of the order one (which
can be, in principle, justified by the need to create fermionic
particles from the vacuum after inflations), we get an estimate
\beq
g \,\sim\, m_f,
\qquad
\tau \,\sim\, m^3_f,
\qquad
f \,\sim\, m_f.
\label{gtauf}
\eeq
For the heaviest particle of the Minimal Standard Model, the
top-quark, we have $m_f = m_t = 175\,GeV$. On the top of that
we can use the value of Hubble parameter during inflation,
\beq
H_{inf} = 10^{11} - 10^{13} GeV,
\quad
\mbox{with}
\quad
|R| \sim  H_{inf}^2,
\mbox{\quad}
\label{gtauf 1}
\eeq
and this gives the estimates
\beq
\sqrt{\frac{6 \xi R}{\la}} \,\sim \, 10^{20}GeV.
\label{gtauf 2}
\eeq
It is easy to see that our ``optimistic'' estimates lead to the situation
that the magnitudes of the non-polynomial terms in the second line
of (\ref{inf5}) are about 3-4 orders of magnitude smaller that the
main  $R^2$-term in the first line. The important consequences
of this fact are that \ {\it i)} The most relevant odd parameter
is $g$, since its contribution is many orders of magnitude greater
that the ones of of $\tau$ and even $f$.
\ {\it ii)} The treatment of the non-polynomial in $R$ terms as
small additions to the main even part, and consequently the same
treatment for the odd terms in the potential compared to the even
terms, is justified.
\ {\it iii)} The interaction of the inflaton with much heavier fermions,
such as the ones of supersymmetric GUT's may lead to the real
trouble with the odd terms in inflation. In this case the
non-polynomial terms become dominating and this implies the
conflict with the observational data, e.g. with the ones of Planck
\cite{Planck}. This problem can be of  course solved by imposing
the small Yukawa couplings for these fermions, that can be seen as
a general restriction for the  inflaton-based models. It is interesting
that we arrived at this conclusion just by requiring the consistency of
the theory of inflaton at the quantum level.
%% red
%%  \textcolor{red}{\LARGE \textbullet}

%%%%%%%%%%%%%%%%%%%%%%%%%%%%
%%%%%%%%%%%%%%%%%%%%%%%%%%%%
\section{Conclusions and Perspectives}
\label{sect5}

As we have seen in the previous sections, in the model with sterile
scalar interacting to fermions there is no symmetry protection from
the terms which are odd in the field, as a result these odd terms are
necessary for the renormalizability of the theory. Similar conclusion
has been done recently in \cite{TomsJHEP,TomsPRD}, but
we were trying to make renormalization in a consistent way that
requires including odd terms into the classical potential.

As far as we include these terms into the classical action, the induced
action of gravity, including both cosmological and inverse Newton
constants, starts to depend on the new terms. And the most
dramatic effect is that the induced action (\r{4}) gains the non-local
contributions with the very small mass (\r{3})  in the Green function.
This illustrated the effect which the sterile scalar (such as, e.g.,
quintessence)  can produce on the gravitational action. It would be
interesting to explore the effect of these non-local terms at the
phenomenological level.

%% red
%% {\LARGE \textbullet}
An interesting consequence can be met by requiring the quantum
consistency of the inflaton coupled to fermions. The simple mapping
to the $F(R)$ models in (\ref{inf5}) shows that the odd terms in
the scalar potential produce the non-polynomial terms in the
gravitational action, with several relevant consequences. In
particular, there may be essential restrictions on the Yukawa
interaction of the inflaton to the heavy fermions beyond the
Standard Model, coming from the Planck data \cite{Planck}.
%% red
%% \textcolor{red}{\LARGE \textbullet}

%%%%%%%%%%%%%%%%%%%%%%%%%%%%%
%%%%%%%%%%%%%%%%%%%%%%%%%%%%%
\section*{Acknowledgments}

The work of I.Sh. was partially supported by Conselho Nacional de
Desenvolvimento Cient\'{i}fico e Tecnol\'{o}gico - CNPq,
303893/2014-1 and Funda\c{c}\~{a}o de Amparo
\`a Pesquisa de Minas Gerais - FAPEMIG, APQ-01205-16.
A.,J.,V. are grateful to Coordena\c{c}\~ao de Aperfei\c{c}oamento de
Pessoal de N\'{\i}vel Superior - CAPES  for supporting their Ph.D.
and MSc projects. I.B. is grateful to CAPES for supporting his
long-term visit to UFJF and to the Physics Departament of UFJF for
kind hospitality. Also he thanks the Russian Ministry of Science and
High Education, project No 3.1386.2017 for partial support.

%%%%%%%%%%%%%%%%%%%%%%%%%%%%%%%%
\section*{Appendix}

The intermediate expressions leading to (\ref{Gadiv}) are
\beq
&&
\nabla_{\mu} \hat{h}^{\mu} = \left(
\begin{array}{cc}
0 & \frac{i}{2} h \nabla_{\mu} \bar{\Psi}_j \gamma^{\mu} \\
0 & \frac{i}{2} h \nabla_{\mu} \ph \gamma^{\mu} \delta^{ij}
\end{array}
\right)
\,,
\,\,
\hat{h}_{\mu} \hat{h}^{\mu} =
\left(
\begin{array}{cc}
0 & - h^{2} \bar{\Psi}_k \ph \\
0 & - h^{2} \ph^{2} \delta^{ik}
\end{array}
\right)
\nn
\\
&& \mathrm{and}
\qquad
\hat{h}_{\mu} \hat{h}_{\nu} = \left(
\begin{array}{cc}
0 & -\frac{1}{4} h^{2} \bar{\Psi}_k \ph \gamma_{\mu} \gamma_{\nu} \\
0 & -\frac{1}{4} h^{2} \ph^{2} \gamma_{\mu} \gamma_{\nu} \delta^{ik}
\end{array}
\right).
\eeq
Thus, we arrive at the expressions
\beq
\hat{P} = \left(
\begin{array}{cc}
\frac{\la \ph^2}{2} + g \ph  + m^2 - \left(\xi - \frac{1}{6}\right) R
& hM \bar{\Psi}_k
- \frac{i}{2} h (\nabla_{\mu} \bar{\Psi}_k) \gamma^{\mu}
+ h^{2} \bar{\Psi}_k \ph
\\
2h \Psi_i & \Big [ M^{2} -\frac{1}{12} R + hM \ph
- \frac{i}{2} h (\nabla_{\mu} \ph)\gamma^\mu + h^2 \ph ^2\Big] \delta^{ik}
\end{array}
\right)
\nn
\eeq
and
\beq
\hat{S}_{\mu \nu} = \left(
\begin{array}{cc}
0 & -\frac{i}{2} h \left( (\nabla_{\mu} \bar{\Psi}_k) \gamma_{\nu}
- (\nabla_{\nu} \bar{\Psi}_k) \gamma_{\mu} \right)
+ \frac{1}{4} h^{2} \bar{\Psi}_k \ph \left[ \gamma_{\mu}, \gamma_{\nu}
\right]
\\
0 & \Big[ \frac{1}{4}R_{\mu \nu \alpha \beta}\gamma^\alpha \ga^\be
- \frac{i}{2} h \left( (\nabla_{\mu} \ph) \gamma_{\nu}
- (\nabla_{\nu} \ph) \gamma_{\mu} \right)
+ \frac{1}{4} h^{2} \ph^{2}
\left[ \gamma_{\mu}, \gamma_{\nu} \right] \Big] \delta ^{ik}
\end{array}\right).
\eeq

%%%%%%%%%%%%%%%%%%%%%%%%%%%%%%%
%%%%%%%%%%%%%%%%%%%%%%%%%%%%%%%
%%%%%%%%%%%%%%%%%%%%%%%%%%%%%%%

%%%%%%%%%%%%%%%%%%%%%%%%%%%

\begin{thebibliography}{99}
%

\bibitem{BuchSha-90}  I.L.~Buchbinder and I.L.~Shapiro,
{\it On the renormalization group equations in curved space-time
with torsion,}
Class. Quant. Grav.  {\bf 7} (1990) 1197.
%%  doi:10.1088/0264-9381/7/7/015

\bibitem{book} I.L. Buchbinder, S.D. Odintsov and I.L. Shapiro,
{\it Effective Action in Quantum Gravity}
(IOP Publishing, Bristol, 1992).

\bibitem{TomsJHEP} D.J.~Toms,
{\it Effective action for the Yukawa model in curved spacetime,}
JHEP {\bf 1805} (2018) 139
%%%  doi:10.1007/JHEP05(2018)139
arXiv:1804.08350.

\bibitem{TomsPRD}  D.J.~Toms,
{\it Gauged Yukawa model in curved spacetime,}
Phys. Rev. {\bf D98}  (2018) 025015,
%%  doi:10.1103/PhysRevD.98.025015
arXiv:1805.01700.

\bibitem{BuchOd84}
I.L. Buchbinder and S.D. Odintsov,
{\it Effective Potential In A Curved Space-time,}
Sov. Phys. J. {\bf 27} (1984) 554.
%%  doi:10.1007/BF00897445}
%% Class. Quant. Grav. {\bf 2} (1985) 721;

\bibitem{BW}  I.L. Buchbinder and J.J. Wolfengaut,
{\it Renormalization Group Equations and Effective Action
in Curved Space-time,} Class. Quant. Grav. {\bf 5} (1988) 1127.

\bibitem{ColeWein} S.R. Coleman and E.J. Weinberg,
{\it Radiative Corrections as the Origin of Spontaneous
Symmetry Breaking,}
Phys. Rev. D7 (1973) 1888.

%  \bibitem{buod-84} I.L. Buchbinder and S.D. Odintsov,
% {\it  Effective Potential In A Curved Space-Time,}
% Sov. Phys. J. 27 (1984) 554; Class. Quant. Grav. 2 (1985) 721.

\bibitem{CorPot} F.~Sobreira, B.J.~Ribeiro and I.L.~Shapiro,
{\it Effective potential in curved space and cut-off regularizations,}
Phys. Lett. {\bf B705} (2011) 273,
%%%    doi:10.1016/j.physletb.2011.10.016
arXiv:1107.2262.

\bibitem{BunchParker79} T.S. Bunch and L. Parker,
{\it Feynman Propagator in Curved Space-Time: A Momentum
Space Representation.}
Phys. Rev. {\bf D20} (1979) 2499.

\bibitem{ParkerToms}   L.~Parker and D.J.~Toms,
{\it Renormalization Group Analysis of Grand Unified Theories in
Curved Space-time,} Phys. Rev. {\bf D29} (1984) 1584.
%%%  doi:10.1103/PhysRevD.29.1584

\bibitem{BuSh-HDQG} I.L.~Buchbinder and I.L.~Shapiro,
{\it Gravitational Interaction Effect on Behavior of the Yukawa and
Scalar Effective Coupling Constants. (In Russian),}
Yad. Fiz.  {\bf 44} (1986) 1033;
\\
I.L.~Buchbinder, O.K.~Kalashnikov, I.L.~Shapiro, V.B.~Vologodsky
and J.J.~Wolfengaut,
{\it The Stability of Asymptotic Freedom in Grand Unified Models
Coupled to $R^{2}$ Gravity,}
Phys. Lett. {\bf B216} (1989) 127;
%% doi:10.1016/0370-2693(89)91381-6
\\
I.L. Shapiro, {\it   Asymptotical behavior of effective Yukawa
coupling constants in quantum $R^2$  -gravity,}
Class. Quant. Grav. {\bf 6} (1989)1197.

\bibitem{Agrav} A.~Salvio and A.~Strumia,
{\it Agravity,}   JHEP {\bf 1406} (2014) 080,
%%  doi:10.1007/JHEP06(2014)080
arXiv:1403.4226.

\bibitem{birdav} N.D. Birell and P.C.W. Davies, {\it Quantum
Fields in Curved Space} (Cambridge University Press, Cambridge,
1982).

\bibitem{BV}
A.O. Barvinsky, G.A. Vilovisky, {\it The generalized Schwinger-Dewitt technique
in gauge theories and quantum gravity,}  Phys.Repts. {\bf 119} (1985) 1.

\bibitem{danteYuk}
G.~de Berredo-Peixoto, D.~D.~Pereira and I.~L.~Shapiro,
{\it Universality and ambiguity in fermionic effective actions,}
Phys. Rev. D {\bf D85} (2012) 064025,
%%%  doi:10.1103/PhysRevD.85.064025
arXiv:1201.2649.

\bibitem{GBP+Gorbar}
G.~de Berredo-Peixoto, E.V.~Gorbar and I.L.~Shapiro,
{\it On the renormalization group for the interacting massive
scalar field theory in curved space,}
Class. Quant. Grav.  {\bf 21} (2004) 2281,
%%  doi:10.1088/0264-9381/21/9/005
hep-th/0311229.

\bibitem{Buch84} I.L. Buchbinder,
{\it On Renormalization Group Equations in Curved Space-Time,}
Theoret. Math. Phys. {\bf 61} (1984) 1215 (Teoret.mat.fiz. {\bf 61} (1984) 393).

\bibitem{Toms83} D.J. Toms,
{\it The Effective Action And The Renormalization Group
Equation In Curved Space-Time,}
Phys. Lett. B126 (1983) 37.

\bibitem{sponta} E.V. Gorbar and I.L. Shapiro,
{\it Renormalization Group and Decoupling in Curved Space:
III. $\,$ The Case of Spontaneous Symmetry Breaking, }
JHEP {\bf 02} (2004) 060,
hep-ph/0311190.

\bibitem{Tmn-ABL} M.~Asorey, P.~M.~Lavrov, B.~J.~Ribeiro
and I.L.~Shapiro,
{\it Vacuum stress-tensor in SSB theories,}
Phys. Rev. {\bf D85} (2012) 104001,
%%  doi:10.1103/PhysRevD.85.104001
arXiv:1202.4235.

\bibitem{GBP}  G.~De Berredo-Peixoto,
{\it A Note on the heat kernel method applied to fermions,}
Mod. Phys. Lett.  {\bf A16} (2001) 2463,
%%  doi:10.1142/S0217732301005965
hep-th/0108223.

\bibitem{Omar-FF} S.A.~Franchino-Vi\~nas, T.~de Paula Netto,
I.L.~Shapiro and O.~Zanusso,
{\it Form factors and decoupling of matter fields in four-dimensional
gravity,}
Phys. Lett. {\bf B790} (2019) 229,
%%  doi:10.1016/j.physletb.2019.01.021
arXiv:1812.00460;
\\
S.~A.~Franchino-Vi\~nas, T.~de Paula Netto and O.~Zanusso,
{\it Vacuum effective actions and mass-dependent renormalization
in curved space,} arXiv:1902.03167.

\bibitem{star} A.A. Starobinski, {\it A new type of isotropic
cosmological models without singularity.} Phys.Lett. {\bf B91} (1980) 99.

\bibitem{star83} A.A.~Starobinsky,
{\it 	The perturbation spectrum evolving from a nonsingular initially
de-Sitter cosmology and the microwave background anisotropy,}
Sov. Astron. Lett. {\bf 9} (1983)  302.

\bibitem{Huang}
Q.G.~Huang, {\it A polynomial $f(R)$ inflation model,}
JCAP {\bf 1402} (2014) 035,
\\
%% doi:10.1088/1475-7516/2014/02/035
arXiv:1309.3514.

\bibitem{AR-RbR}
A.R.R.~Castellanos, F.~Sobreira, I.L.~Shapiro and A.A.~Starobinsky,
{\it On higher derivative corrections to the $R+R^2$ inflationary
model,}
JCAP {\bf 1812} (2018) 007,
%%  doi:10.1088/1475-7516/2018/12/007
arXiv:1810.07787.

\bibitem{StabInstab}
T.d.P.~Netto, A.M.~Pelinson, I.L.~Shapiro and A.A.~Starobinsky,
{\it From stable to unstable anomaly-induced inflation,}
Eur. Phys. J. {\bf C76} (2016) 544,
%%  doi:10.1140/epjc/s10052-016-4390-4
arXiv:1509.08882.

\bibitem{Shaposh} F.L. Bezrukov and M. Shaposhnikov,
{\it The Standard Model Higgs boson as the inflaton,}
Phys. Lett. {\bf B659} (2008) 703,
%%  doi:10.1016/j.physletb.2007.11.072
arXiv:0710.3755.

\bibitem{Planck}
P.~A.~R.~Ade {\it et al.} [Planck Collaboration],
{\it Planck 2015 results. XX. Constraints on inflation,}
Astron. Astrophys.  {\bf 594} (2016) A20,
%%%  doi:10.1051/0004-6361/201525898
arXiv:1502.02114.

%
%%%%%%%%%%%%%%%%%%%%%%%%%%%
\end{thebibliography}
\end{document}